\documentclass[10pt, conference, a4paper, oneside, twocolumn]{IEEEtran}
\usepackage{graphicx}
\usepackage{amsmath}
\usepackage{amssymb}
\usepackage{mathrsfs}
\usepackage{stfloats}
\usepackage{dsfont}
\usepackage{setspace}
\begin{document}
%
%
%
%
%
%
\title{Diversity, Coding, and Multiplexing Trade--Off of Network--Coded Cooperative Wireless Networks}
\author{\authorblockN{Michela Iezzi$^{(1)}$, Marco Di Renzo$^{(2)}$, Fabio Graziosi$^{(1)}$}
\authorblockA{\footnotesize $^{(1)}$ University of L'Aquila, College of Engineering\\
Department of Electrical and Information Engineering (DIEI), Center of Excellence of Research DEWS\\
Via G. Gronchi 18, Nucleo Industriale di Pile, 67100 L'Aquila, Italy\\
$^{(2)}$ L2S, UMR 8506 CNRS -- SUPELEC -- Univ Paris--Sud\\
Laboratory of Signals and Systems (L2S), French National Center for Scientific Research (CNRS)\\
\'Ecole Sup\'erieure d'\'Electricit\'e (SUP\'ELEC), University of Paris--Sud XI (UPS)\\
3 rue Joliot--Curie, 91192 Gif--sur--Yvette (Paris), France\\
E--Mail: marco.direnzo@lss.supelec.fr, \{michela.iezzi, fabio.graziosi\}@univaq.it} }
\maketitle
\begin{abstract}
In this paper, we study the performance of network--coded cooperative diversity systems with practical communication constraints. More
specifically, we investigate the interplay between diversity, coding, and multiplexing gain when the relay nodes do not act as dedicated
repeaters, which only forward data packets transmitted by the sources, but they attempt to pursue their own interest by forwarding packets
which contain a network--coded version of received and their own data. We provide a very accurate analysis of the Average Bit Error Probability
(ABEP) for two network topologies with three and four nodes, when practical communication constraints, \emph{i.e.}, erroneous decoding at the
relays and fading over all the wireless links, are taken into account. Furthermore, diversity and coding gain are studied, and advantages and
disadvantages of cooperation and binary Network Coding (NC) are highlighted. Our results show that the throughput increase introduced by NC is
offset by a loss of diversity and coding gain. It is shown that there is neither a coding nor a diversity gain for the source node when the
relays forward a network--coded version of received and their own data. Compared to other results available in the literature, the conclusion
is that binary NC seems to be more useful when the relay nodes act only on behalf of the source nodes, and do not mix their own packets to the
received ones. Analytical derivation and findings are substantiated through extensive Monte Carlo simulations.
\end{abstract}
%
%
%
%
%
\section{Introduction} \label{Intro}
\PARstart{C}{ooperative}/multi--hop networking has recently emerged as a strong candidate technology for many future wireless applications
\cite{Nosratinia}, \cite{Laneman}. The basic premise of cooperative/multi--hop communications is to achieve and to exploit the benefits of spatial
diversity without requiring each mobile node to be equipped with co--located multiple antennas. On the contrary, each mobile node becomes part
of a large distributed array and shares its single--antenna (as well as hardware, processing, and energy resources) to help other nodes of the
network to achieve better performance/coverage. However, the efficient exploitation of cooperative/multi--hop networking is faced by the
following challenges \cite{Krikidis}, \cite{Zorzi}: i) due to practical considerations, such as the half--duplex constraint or to avoid
interference caused by simultaneous transmissions, distributed cooperation needs extra bandwidth resources (\emph{e.g.}, time slots or
frequencies), which might result in a loss of system throughput; ii) relay nodes are forced to use their own resources to forward the packets
of other nodes, usually without receiving any rewards, except for the fact that the whole system can become more efficient; and iii) in
classical cooperative protocols, the relay nodes that perform a retransmission on behalf of other nodes must delay their own frames, which has
an impact on the latency of the network.

To overcome these limitations, a new technology named Network Coding (NC) has recently been introduced to improve the network performance
\cite{Ahlswede}--\cite{Katti_PhD}. NC can be broadly defined as an advanced routing or encoding mechanism at the network layer, which allows
network nodes not only to forward but also to process incoming data packets. Different forms of NC exist in the literature, \emph{e.g.},
algebraic NC, physical--layer NC, and Multiple--Input--Multiple--Output (MIMO--) NC, which offer a different trade--off between achievable
performance and implementation complexity. The interested reader might consult \cite{Zorzi} for a recent survey and comparison of these
methods. The common feature of all NC approaches is that the network throughput is improved by allowing some network nodes to combine many
incoming packets, which, after being mixed, need a single wireless resource (\emph{e.g.}, a time slot or a frequency) for their transmission.
Thus, NC is considered a potential and effective enabler to recover the throughput loss experienced by cooperative/multi--hop networking
\cite{Krikidis}. Theory and experiments have shown that network--coded cooperative/multi--hop systems can be extremely useful for wireless
networks with disruptive channel and connectivity conditions \cite{Gerla}, \cite{Katti_PhD}.

The performance of cooperative/multi--hop networks has been studied extensively during the last years, see, \emph{e.g.},
\cite{Ribeiro}--\cite{MDR_TCOMFeb2010}, and many important conclusions have been drawn about the achievable diversity and coding gain over
fading channels. On the other hand, the analysis of the performance of cooperative/multi--hop systems with NC is almost unexplored so far. More
specifically, understanding the interplay between the multiplexing gain introduced by NC and the achievable diversity/coding gain introduced by
cooperation is an open and challenging research problem, especially when practical communication constraints (erroneous decoding and fading)
are taken into account \cite{MDR_Springer2010}--\cite{AlHabian2011}. Some recent results on this matter are
\cite{Cano}--\cite{Iezzi_TIT2011}. In particular, \cite{Nasri} and \cite{Iezzi_GLOBECOM2011} have recently provided an accurate and
closed--form analysis of network--coded cooperative/multi--hop systems by estimating both diversity and coding gain with realistic
source--to--relay links. These papers have highlighted, for some network topologies and encoding schemes, the potential benefits of NC to
recover the throughput loss of cooperative/multi--hop networking.

However, the analysis in \cite{Nasri} and \cite{Iezzi_GLOBECOM2011} considers the classical scenario where some network nodes (\emph{i.e.}, the
relays) operate only on behalf of other network nodes (\emph{i.e.}, the sources) when forwarding data to a given destination. In other words,
the relays are dedicated network elements with no data to transmit and, thus, they receive no direct reward from cooperation. In this paper, we are
interested in studying the interplay between diversity, coding, and multiplexing gain of network--coded cooperative/multi--hop wireless
networks when the relays have their own data packets to be transmitted to a common destination, and exploit NC to transmit them along with the
packets that have to be relayed on behalf of the sources. This way, the relays can help the sources without the need to: i) delay the
transmission of their own data packets; and ii) use specific resources (energy and processing) to forward the packets of the sources. Thus, NC
can potentially avoids throughput and energy loss. However, it is not clear whether performing NC at the relay nodes entail any performance
(\emph{i.e.}, diversity or coding gain) loss with respect to classical cooperative diversity. The main aim of this paper is to shed lights on
this matter, and to highlight the fundamental diversity, coding, and multiplexing trade--off with realistic communication constraints and
binary NC at the relays. To this end, two network topologies are considered with 3 (1 source, 1 relay, 1 destination) and 4 nodes (1 source, 2
relays, 1 destination), and the end--to--end Average Bit Error Probability (ABEP) over independent but non--identically distributed (i.n.i.d)
Rayleigh fading channels is computed in closed--form. Our results highlight that the throughput increase introduced by NC is offset by a loss
of the diversity gain. More specifically, it is shown that, when the relays forward a network--coded version of received and their own data
packets, there is neither a coding nor a diversity gain for the source. Compared to other results available in the literature \cite{Nasri},
\cite{Iezzi_GLOBECOM2011}, the conclusion is that binary NC seems to be more useful when the relays act on behalf of the sources only, and do
not mix their own packets to the received ones.

The remainder of this paper is organized as follows. In Section \ref{SystemModel}, system model and problem statement are summarized. In Section
\ref{Framework}, the analytical framework to compute the ABEP is described. In Section \ref{Comparison}, the achievable diversity, coding, and
multiplexing gain of various schemes with and without NC are analyzed and compared. In Section \ref{Results}, some numerical results are shown.
Finally, Section \ref{Conclusion} concludes this paper.
\begin{figure}[!t]
\centering\includegraphics [width=0.40\columnwidth] {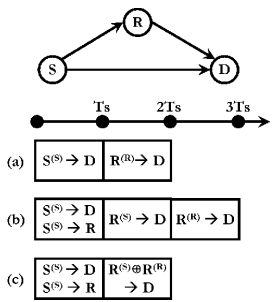} \caption{\scriptsize 1--source ($S$), 1--relay ($R$),
1--destination ($D$) network topology. Nodes $S$ and $R$ have data packets to transmit to $D$. $X^{\left( Y \right)}  \to Z$ denotes that node
$X$ processes/manipulates the data packet of node $Y$ to forward it to node $Z$. Scenarios: (a) non--cooperative; (b) cooperative ($R$ acts as
a relay for $S$); and (c) network--coded cooperative ($R$ acts as a relay for $S$ and at the same time transmits its own data to $D$).}
\label{Fig_1} \vspace{-0.25cm}
\end{figure}
\begin{figure}[!t]
\centering\includegraphics [width=0.70\columnwidth] {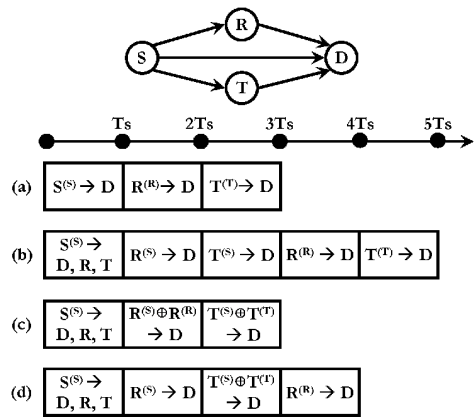} \caption{\scriptsize 1--source ($S$), 2--relay ($R$ and
$T$), 1--destination ($D$) network topology. Nodes $S$, $R$, and $T$ have data packets to transmit to $D$. Notation: i) $X^{\left( Y \right)}
\to Z$ denotes that node $X$ processes/manipulates the data packet of node $Y$ to forward it to node $Z$. Scenarios: (a) non--cooperative; (b)
cooperative ($R$ and $T$ act as relays for $S$); (c) network--coded cooperative ($R$ and $T$ act as relays for $S$ and at the same time
transmit their own data to $D$); (d) hybrid network--coded cooperative ($R$ acts only as a relay for $S$, while $T$ acts as a relay for $S$ and
at the same time transmits its own data to $D$).} \label{Fig_2} \vspace{-0.25cm}
\end{figure}
\begin{figure*}[!t]
\setcounter{equation}{1}
\begin{equation} \scriptsize
\label{Eq_2} \hspace{-0.3cm} \left[ {\hat b_S^{\left( D \right)} ,\hat b_T^{\left( D \right)} } \right] = \mathop {\arg \min
}\limits_{\scriptstyle \tilde b_S \in \left\{ {0,1} \right\} \hfill \atop
  \scriptstyle \tilde b_T  \in \left\{ {0,1} \right\} \hfill} \left\{ {\underbrace {\frac{{\left| {y_{SD}  - \sqrt {E_m } h_{SD} \left( {1 - 2\tilde b_S } \right)} \right|^2 }}{{N_0 }} + \lambda _R \frac{{\left| {y_{RD}  - \sqrt {{{E_m } \mathord{\left/
 {\vphantom {{E_m } 2}} \right.
 \kern-\nulldelimiterspace} 2}} h_{RD} \left( {1 - 2\tilde b_S } \right)} \right|^2 }}{{N_0 }} + \lambda _T \frac{{\left| {y_{TD}  - \sqrt {E_m } h_{TD} \left[ {1 - 2\left( {\tilde b_S  \oplus \tilde b_T } \right)} \right]} \right|^2 }}{{N_0 }}}_{\Lambda \left( {\tilde b_S ,\tilde b_T ;b_S ,b_T ,\hat b_S^{\left( R \right)} ,\hat b_S^{\left( T \right)} } \right)} } \right\}
\vspace{-5pt}
\end{equation}
\normalsize \hrulefill \vspace*{-5pt}
\end{figure*}
\section{System Model and Problem Statement} \label{SystemModel}
We study two cooperative network topologies with three and four nodes, as shown in Fig. \ref{Fig_1} and Fig. \ref{Fig_2}, respectively. We
consider a Time--Division--Multiple--Access (TDMA) protocol, where all transmissions take place in non--overlapping time--slots ($T_S$ denotes
the duration of a time--slot). Also, we assume the half--duplex constraint, \emph{i.e.}, nodes cannot transmit and receive at the same time
\cite{Krikidis}. Furthermore, we analyze the MIMO--NC approach, where network decoding and demodulation at the final destination are jointly
performed at the physical layer, which results in a cross--layer decoding algorithm \cite{Zorzi}. For analytical tractability, we assume that
each node uses uncoded Binary Phase Shift Keying (BPSK) modulation. In those scenarios where NC is exploited, we consider binary NC (exclusive
OR denoted by $\oplus$) as this provides a low--complexity design of the relays. Each wireless channel is assumed to experience Rayleigh
fading. More specifically, the fading coefficient between two generic nodes $X$ and $Y$ is denoted by $h_{XY}$, and it is assumed to be a
circular symmetric complex Gaussian Random Variable (RV) with zero mean and variance $\sigma _{XY}^2$ per dimension. Fading
over different links is assumed to be i.n.i.d to account for different propagation distances and shadowing effects. The noise at the input of
node $Y$ and related to the transmission from node $X$ to node $Y$ is denoted by $n_{XY}$, and it is assumed to be complex Additive White
Gaussian (AWG) with variance ${{N_0 } \mathord{\left/{\vphantom {{N_0 } 2}} \right. \kern-\nulldelimiterspace} 2}$ per dimension. Finally, $n_{XY}$ at different time--slots or at the input of different nodes are assumed to be independent and identically distributed (i.i.d.).
\subsection{Problem Statement} \label{ProblemStatement}
The main objective of this paper is to understand the performance vs. throughput trade--off provided by NC over fading channels. To be more
specific, let us consider the 3--node scenario in Fig. \ref{Fig_1}. Similar comments apply to the 4--node scenario in Fig. \ref{Fig_2}. We have
two nodes ($S$ and $R$), which have data to transmit to node $D$. In Scenario (a), both nodes perform their transmission to $D$ in a selfish
mode, \emph{i.e.}, no cooperation. In Scenario (b), node $R$ is willing to help node $S$ to forward the overheard packet to node $D$. In this
case, node $S$ acts as a ``golden user'', and node $R$ delays the transmission of its own data packet to help node $S$ first. In this case,
node $S$ can take advantage of cooperation to improve its performance. However, node $R$ has to share its transmission energy with node $S$,
and it must delay its own transmission: this is the price of cooperation. In Scenario (c), node $R$ uses NC to avoid the limitations just
mentioned. By using NC, node $R$ can avoid to delay its own packet, and it can transmit a coded (XOR) version of overheard packet from node $S$
and its own packet. The gain is twofold: i) no transmission delay; and ii) no need to share transmission energy with node $S$. In this
case, the overall transmission can be completed in two time--slots rather than in three time--slots as in Scenario (b). Thus the network
throughput increases.

The fundamental questions we want to address in this paper are: i) \emph{Is there any performance (diversity/coding gain) loss, with respect to
selfish and cooperative scenarios, for this throughput gain}?; and ii) \emph{In case of performance loss, is this only due to erroneous
decoding at node $R$ or is this related to NC operations too}? Our closed--form asymptotic analysis will provide a clear answer to both
questions. Similar questions hold for Fig. \ref{Fig_2} as well, where we can see that, depending on the level of cooperation and NC, the
throughput of the network, \emph{i.e.}, the number of time--slots, is different.

Due to space limitations, we are unable to provide a step--by--step analysis and derivation for all the scenarios shown in Fig. \ref{Fig_1} and
Fig. \ref{Fig_2}. However, the analytical development is very similar for all of them. Thus, for ease of exposition and clarity, we have
decided to focus our attention on a scenario only. We have chosen Scenario (d) in Fig. \ref{Fig_2}, as it is the most general one. So, in the
remainder of this paper only this scenario will be analyzed analytically. However, in Section \ref{Comparison} we will summarize the final
expression of the ABEP for all the scenarios in Fig. \ref{Fig_1} and Fig. \ref{Fig_2}, and we will compare achievable performance and throughput of all of them.
\subsection{Signal Model} \label{SignalModel}
Let us consider Scenario (d) in Fig. \ref{Fig_2}. During the first time--slot, node $S$ broadcasts a BPSK modulated bit, $x_{S} = \sqrt {E_m }
\left( {1 - 2b_{S} } \right)$, where $E_m$ is the average transmitted energy and $b_{S}  \in \left\{ {0,1} \right\}$ is the bit emitted by $S$.
The signals received at nodes $R$, $T$, and $D$ are given by $y_{S X}  = h_{S X } x_{S }  + n_{S X }$, where $X=R$, $X=T$, and $X=D$,
respectively. Similar to \cite{Nasri}, \cite{Iezzi_ICC2011}, \cite{Iezzi_GLOBECOM2011}, the intermediate nodes $R$ and $T$ demodulate the
received bit by using conventional Maximum--Likelihood (ML--) optimum decoding:
\setcounter{equation}{0}
\begin{equation} \scriptsize
\label{Eq_1} \hat b_S^{\left( X \right)}  = \mathop {\arg \min }\limits_{\tilde b_S  \in \left\{ {0,1} \right\}} \left\{ {\left| {y_{SX}  -
\sqrt {E_m } h_{SX} \left( {1 - 2\tilde b_S } \right)} \right|^2 } \right\}
\end{equation}
\noindent where $X=R$ and $X=T$, and $({\hat{\cdot} })$ and $({\tilde {\cdot}})$ denote detected/estimated and trial bit of the
hypothesis--detection problem, respectively. $\hat b_S^{\left( X \right)}$ is the estimate of $b_S$ at node $X$.

During the second time--slot, node $R$ remodulates and forwards its estimate of $b_S$, \emph{i.e.}, $\hat b_S^{\left( R \right)}$, to node $D$.
The transmitted bit is $x_R  = \sqrt {{{E_m } \mathord{\left/ {\vphantom {{E_m } 2}} \right. \kern-\nulldelimiterspace} 2}} \left( {1 - 2\hat
b_S^{\left( R \right)} } \right)$. Let us note that node $R$ uses only half of its available energy to forward $\hat b_S^{\left( R \right)}$ on
behalf of node $S$, as it needs half energy to transmit its own data during the fourth time--slot. This allows us to consider a total energy
constraint, and it guarantees a fair comparison among the scenarios. Similar considerations apply to all the scenarios shown in Fig. \ref{Fig_1}
and Fig. \ref{Fig_2}. The signal received at node $D$ is $y_{RD}  = h_{RD} x_R  + n_{RD}$.

During the third time--slot, node $T$ performs similar operations as node $R$ in the second time--slot. However, node $T$ applies binary NC to
avoid to use two time--slots to help nodes $S$ and to transmit its own data. More specifically, the bit transmitted by node $T$ is $x_T = \sqrt
{E_m } \left[ {1 - 2\left( {\hat b_S^{\left( T \right)}  \oplus b_T } \right)} \right]$, where $b_T$ is the bit that $T$ wants to transmit to
node $D$. Unlike node $R$, node $T$ uses full transmission energy, since, with the help of NC, it does not need an extra time--slot to forward
its own data. The signal received at $D$ is $y_{TD}  = h_{TD} x_T  + n_{TD}$.

Finally, let us note that the fourth time--slot is not of interest in the detection process, as the bit transmitted in this time--slot is
independent of all the others. So, it can be demodulated without considering previous received bits. However, the need of this time--slot to
complete the overall communication is important to assess the network throughput of the system.
\subsection{Detection at Node $D$} \label{Receiver}
Upon reception of signals ${y_{SD}}$, ${y_{RD}}$, and ${y_{TD}}$ in time--slot one, two, and three, respectively, node $D$ can perform joint
demodulation of $b_S$ and $b_T$. As mentioned above, $b_R$ is treated independently as the related packet is independent of the others. To
avoid the analytical intractability and implementation complexity of the ML--optimum demodulator, we consider the sub--optimal, but
asymptotically--tight (for high Signal--to--Noise--Ratio, SNR), Cooperative Maximum Ratio Combining (C--MRC) detector shown in (\ref{Eq_2}) on
top of this page \cite{Nasri}, \cite{GiannakisLaneman}, where: i) $\lambda _R  = {{\min \left\{ {\gamma _{SR} ,\gamma _{RD} } \right\}}
\mathord{\left/ {\vphantom {{\min \left\{ {\gamma _{SR} ,\gamma _{RD} } \right\}} {\gamma _{RD} }}} \right. \kern-\nulldelimiterspace} {\gamma
_{RD} }}$ and $\lambda _T  = {{\min \left\{ {\gamma _{ST} ,\gamma _{TD} } \right\}} \mathord{\left/ {\vphantom {{\min \left\{ {\gamma _{ST}
,\gamma _{TD} } \right\}} {\gamma _{TD} }}} \right. \kern-\nulldelimiterspace} {\gamma _{TD} }}$ account for the reliability of the
$S$--to--$R$ and $S$--to--$T$ links, respectively; and ii) $\gamma _{XY}  = \left| {h_{XY} } \right|^2 \left( {{{E_m } \mathord{\left/
{\vphantom {{E_m } {N_0 }}} \right. \kern-\nulldelimiterspace} {N_0 }}} \right)$ with $X$ and $Y$ being two generic nodes of the network. The
derivation of (\ref{Eq_2}) follows the same arguments as in \cite{Nasri}, \cite{GiannakisLaneman}, and it is here omitted to avoid repetitions.
\begin{figure*}[!t]
\setcounter{equation}{3}
\begin{equation} \scriptsize
\label{Eq_4}
\begin{split}
  & \hspace{-0.5cm} {\rm{APEP}}\left( {{\bf{c}} \to {\bf{\tilde c}}} \right) = \Pr \left\{ {\Lambda _{{\bf{\tilde c}}}  < \Lambda _{\bf{c}} } \right\} = \Pr \left\{ {\Delta _{{\bf{c}},{\bf{\tilde c}}}  = \Lambda _{{\bf{\tilde c}}}  - \Lambda _{\bf{c}}  < 0} \right\} \\
 & \hspace{-0.5cm} \mathop  = \limits^{\left( a \right)} {\rm{E}}_{h_{SR} ,h_{ST} } \left\{ {\Pr \left\{ {\left. {\Delta _{{\bf{c}},{\bf{\tilde c}}}  < 0 } \right|\hat b_S^{\left( R \right)}  = b_S ,\hat b_S^{\left( T \right)}  = b_S } \right\}\Pr \left\{ {\hat b_S^{\left( R \right)}  = b_S ,\hat b_S^{\left( T \right)}  = b_S } \right\}} \right\}
  + {\rm{E}}_{h_{SR} ,h_{ST} } \left\{ {\Pr \left\{ {\left. {\Delta _{{\bf{c}},{\bf{\tilde c}}}  < 0 } \right|\hat b_S^{\left( R \right)}  = b_S ,\hat b_S^{\left( T \right)}  \ne b_S } \right\}\Pr \left\{ {\hat b_S^{\left( R \right)}  = b_S ,\hat b_S^{\left( T \right)}  \ne b_S } \right\}} \right\} \\
 & \hspace{-0.5cm} + {\rm{E}}_{h_{SR} ,h_{ST} } \left\{ {\Pr \left\{ {\left. {\Delta _{{\bf{c}},{\bf{\tilde c}}}  < 0 } \right|\hat b_S^{\left( R \right)}  \ne b_S ,\hat b_S^{\left( T \right)}  = b_S } \right\}\Pr \left\{ {\hat b_S^{\left( R \right)}  \ne b_S ,\hat b_S^{\left( T \right)}  = b_S } \right\}} \right\}
  + {\rm{E}}_{h_{SR} ,h_{ST} } \left\{ {\Pr \left\{ {\left. {\Delta _{{\bf{c}},{\bf{\tilde c}}}  < 0 } \right|\hat b_S^{\left( R \right)}  \ne b_S ,\hat b_S^{\left( T \right)}  \ne b_S } \right\}\Pr \left\{ {\hat b_S^{\left( R \right)}  \ne b_S ,\hat b_S^{\left( T \right)}  \ne b_S } \right\}} \right\} \\
 \end{split}
\end{equation}
\normalsize \hrulefill \vspace*{-10pt}
\end{figure*}
\section{Performance Analysis} \label{Framework}
The aim of this section is to estimate the performance of the detector in (\ref{Eq_2}), by providing a closed--form expression of the ABEP for
high--SNR. The ABEP of node $S$ and node $T$, \emph{i.e.}\footnote{$\Pr \left\{  \cdot  \right\}$ denotes probability.}, ${\rm{ABEP}}_S  = \Pr
\left\{ {b_S \ne \hat b_S^{\left( D \right)} } \right\}$ and ${\rm{ABEP}}_T  = \Pr \left\{ {b_T  \ne \hat b_T^{\left( D \right)} } \right\}$,
respectively, can be computed by using the methodology described in \cite[Sec. IV]{Iezzi_ICC2011}. In particular, we have:
\setcounter{equation}{2}
\begin{equation} \scriptsize
\label{Eq_3} {\rm{ABEP}}_X  \le \frac{1}{{{\rm{card}}\left\{ \mathcal{C} \right\}}}\sum\limits_{b_S  = 0}^1 {\sum\limits_{\tilde b_S  = 0}^1
{\sum\limits_{b_T  = 0}^1 {\sum\limits_{\tilde b_T  = 0}^1 {{\rm{APEP}}_X \left( {{\bf{c}} \to {\bf{\tilde c}}} \right)} } } }
\end{equation}
\noindent where: i) ${\rm{APEP}}_X \left( {{\bf{c}} \to {\bf{\tilde c}}} \right){\rm{ = APEP}}\left( {{\bf{c}} \to {\bf{\tilde c}}} \right)\bar
\Delta \left( {b_X , {\tilde b}_X } \right)$; ii) $\mathcal{C} = \left\{ {000,010,111,101} \right\}$ is the codebook of Scenario (d) in Fig.
\ref{Fig_2}, which takes into account forwarding and NC operations performed at nodes $R$ and $T$. The generic element of $\mathcal{C}$ is
${\bf{c}} = \left[ {b_S ,b_S ,b_S \oplus b_T } \right]$; iii) ${\rm{card}}\left\{ \mathcal{C} \right\} = 4$ is the cardinality of
$\mathcal{C}$, \emph{i.e.}, the number of codewords ${\bf{c}}$ in $\mathcal{C}$; iv) ${\rm{APEP}}\left( {{\bf{c}} \to {\bf{\tilde c}}} \right)$
is the Average Pairwise Error Probability (APEP) of the generic pair of codewords ${\bf{c}} = \left[ {c_1 ,c_2 ,c_3 } \right] = \left[ {b_S
,b_S ,b_S  \oplus b_T } \right]$ and ${\bf{\tilde c}} = \left[ {\tilde c_1 ,\tilde c_2 ,\tilde c_3 } \right] = \left[ {\tilde b_S ,\tilde b_S
,\tilde b_S \oplus \tilde b_T } \right]$ of the codebook, \emph{i.e.}, the probability of estimating ${\bf{\tilde c}}$ in (\ref{Eq_2}), when,
instead, ${\bf{c}}$ has actually been transmitted, and ${\bf{c}}$ and ${\bf{\tilde c}}$ are the only two codewords possibly being transmitted;
and v) $\bar \Delta \left(b_X ,\tilde b_X \right) = 1 - \Delta \left( b_X ,\tilde b_X  \right)$, where $\Delta \left( {\cdot,\cdot} \right)$ is
the Kronecker delta function, \emph{i.e.}, $\Delta \left( {b_X ,\tilde b_X } \right) = 1$ if $b_X  = \tilde b_X$ and $\Delta \left( {b_X
,\tilde b_X } \right) = 0$ if $b_X  \ne \tilde b_X$. This function is used to include in the computation of ${\rm{ABEP}}_X$ only those APEPs
which result in an error for the information bit of interest, \emph{i.e.}, $X=S$ or $X=T$ \cite{Iezzi_ICC2011}.
\subsection{Computation of ${\rm{APEP}}\left( {{\bf{c}} \to {\bf{\tilde c}}} \right)$} \label{APEP}
From (\ref{Eq_3}), it follows that that ABEP can be estimated if ${\rm{APEP}}\left( {{\bf{c}} \to {\bf{\tilde c}}} \right)$ is available in
closed--form, where the average is over fading channel statistics and AWGN. In this section, we compute an asymptotically--tight formula for
${\rm{APEP}}\left( {{\bf{c}} \to {\bf{\tilde c}}} \right)$, which is accurate for high--SNR.

From (\ref{Eq_2}), by definition, we have (\ref{Eq_4}) on top of this page, where: i) $\Lambda _{\bf{c}}  = \Lambda \left( {b_S ,b_T ;b_S ,b_T
,\hat b_S^{\left( R \right)} ,\hat b_S^{\left( T \right)} } \right)$ and $\Lambda _{{\bf{\tilde c}}}  = \Lambda \left( {\tilde b_S ,\tilde b_T
;b_S ,b_T ,\hat b_S^{\left( R \right)} ,\hat b_S^{\left( T \right)} } \right)$; ii) ${\rm{E}}_X \left\{  \cdot  \right\}$ is the expectation
operator computed over RV $X$; and iii) $\mathop  = \limits^{\left( a \right)}$ is obtained by using the total probability theorem and by
conditioning upon possible decoding errors at nodes $R$ and $T$ \cite{Proakis}. Since demodulation outcomes at node $R$ and $T$ are
independent, we have: i) {\footnotesize $\Pr \left\{ {\hat b_S^{\left( R \right)} = b_S ,\hat b_S^{\left( T \right)}  = b_S } \right\} = \left[
{1 - Q\left( {\sqrt {2\gamma _{SR} } } \right)} \right]\left[ {1 - Q\left( {\sqrt {2\gamma _{ST} } } \right)} \right]$}; ii) {\footnotesize
$\Pr \left\{ {\hat b_S^{\left( R \right)}  = b_S ,\hat b_S^{\left( T \right)}  \ne b_S } \right\} = \left[ {1 - Q\left( {\sqrt {2\gamma _{SR} }
} \right)} \right]Q\left( {\sqrt {2\gamma _{ST} } } \right)$}; iii) {\footnotesize $\Pr \left\{ {\hat b_S^{\left( R \right)} \ne b_S ,\hat
b_S^{\left( T \right)}  = b_S } \right\} = Q\left( {\sqrt {2\gamma _{SR} } } \right)\left[ {1 - Q\left( {\sqrt {2\gamma _{ST} } } \right)}
\right]$}; and iv) {\footnotesize $\Pr \left\{ {\hat b_S^{\left( R \right)}  \ne b_S ,\hat b_S^{\left( T \right)} \ne b_S } \right\} = Q\left(
{\sqrt {2\gamma _{SR} } } \right)Q\left( {\sqrt {2\gamma _{ST} } } \right)$}, where $Q\left( x \right) = \left( {{1 \mathord{\left/ {\vphantom
{1 {\sqrt {2\pi } }}} \right. \kern-\nulldelimiterspace} {\sqrt {2\pi } }}} \right)\int_x^{ + \infty } {\exp \left( { - {{t^2 } \mathord{\left/
{\vphantom {{t^2 } 2}} \right. \kern-\nulldelimiterspace} 2}} \right)dt}$ is the Q--function and these probabilities are due to using BPSK
modulation \cite{Proakis}. From these expressions, it follows that conditioning upon decoding errors at node $R$ and node $T$ implies
conditioning upon the fading channel gains $h_{SR}$ and $h_{ST}$. This explains the presence of the expectations in (\ref{Eq_4}).
\begin{figure*}[!t]
\setcounter{equation}{6}
\begin{equation} \scriptsize
\label{Eq_7}
\begin{split}
  {\rm{APEP}}^{\left( {\rm{4}} \right)} \left( {{\bf{c}} \to {\bf{\tilde c}}} \right) &= \frac{1}{{2\pi j}}\int\nolimits_{\delta  - j\infty }^{\delta  + j\infty } {{\rm{E}}_{\left\{ {h_{XY} } \right\},\left\{ {n_{XY} } \right\}} \left\{ {\exp \left[ { - s\mathcal{F}\left( {\left\{ {h_{XY} } \right\},\left\{ {\bar n_{XY} } \right\}} \right)} \right]Q\left( {\sqrt {2\gamma _{SR} } } \right)Q\left( {\sqrt {2\gamma _{ST} } } \right)} \right\}\frac{{ds}}{s}}  \\
 &\mathop  = \limits^{\left( a \right)} \frac{1}{{2\pi ^3 j}}\int\nolimits_{\delta  - j\infty }^{\delta  + j\infty } {\int\nolimits_{\rm{0}}^{{\pi  \mathord{\left/
 {\vphantom {\pi  {\rm{2}}}} \right.
 \kern-\nulldelimiterspace} {\rm{2}}}} {\int\nolimits_{\rm{0}}^{{\pi  \mathord{\left/
 {\vphantom {\pi  {\rm{2}}}} \right.
 \kern-\nulldelimiterspace} {\rm{2}}}} {{\rm{E}}_{\left\{ {h_{XY} } \right\},\left\{ {n_{XY} } \right\}} \left\{ {\exp \left[ { - s\mathcal{F}\left( {\left\{ {h_{XY} } \right\},\left\{ {\bar n_{XY} } \right\}} \right)} \right]\exp \left( { - \frac{{\gamma _{SR} }}{{\sin ^2 \left( {\theta _1 } \right)}}} \right)\exp \left( { - \frac{{\gamma _{ST} }}{{\sin ^2 \left( {\theta _2 } \right)}}} \right)} \right\}d\theta _1 d\theta _2 \frac{{ds}}{s}} } }  \\
 &\mathop  = \limits^{\left( b \right)} \frac{1}{{2\pi ^3 j}}\int\nolimits_{\delta  - j\infty }^{\delta  + j\infty } {\int\nolimits_{\rm{0}}^{{\pi  \mathord{\left/
 {\vphantom {\pi  {\rm{2}}}} \right.
 \kern-\nulldelimiterspace} {\rm{2}}}} {\int\nolimits_{\rm{0}}^{{\pi  \mathord{\left/
 {\vphantom {\pi  {\rm{2}}}} \right.
 \kern-\nulldelimiterspace} {\rm{2}}}} {\mathcal{G}\left( {s,\theta _1 ,\theta _2 } \right)d\theta _1 d\theta _2 \frac{{ds}}{s}} } } \mathop  = \limits^{\left( c \right)} \frac{1}{{2\pi ^3 j}}\int\nolimits_{\delta  - j\infty }^{\delta  + j\infty } {\Psi _0 \left( s \right)\left( {\int\nolimits_{\rm{0}}^{{\pi  \mathord{\left/
 {\vphantom {\pi  {\rm{2}}}} \right.
 \kern-\nulldelimiterspace} {\rm{2}}}} {\Psi _1 \left( {s,\theta _1 } \right)d\theta _1 } } \right)\left( {\int\nolimits_{\rm{0}}^{{\pi  \mathord{\left/
 {\vphantom {\pi  {\rm{2}}}} \right.
 \kern-\nulldelimiterspace} {\rm{2}}}} {\Psi _2 \left( {s,\theta _2 } \right)d\theta _2 } } \right)\frac{{ds}}{s}}  \\
 \end{split} \vspace*{-10pt}
\end{equation}
\normalsize \hrulefill \vspace*{-10pt}
\end{figure*}
\begin{figure*}[!t]
\setcounter{equation}{7}
\begin{equation} \scriptsize
\label{Eq_8} \mathcal{F}\left( {\left\{ {\gamma_{XY} } \right\},\left\{ {\bar n_{XY} } \right\}} \right) = \gamma _{SD} d_S^2  + 2\sqrt {\gamma
_{SD} } d_S {\mathop{\rm Re}\nolimits} \left\{ {\bar n_{SD}^ *  } \right\} + \lambda _R \left( {\gamma _{RD} \hat d_R^{\left( {{\rm{nok}}}
\right)} + 2\sqrt {\gamma _{RD} } d_R {\mathop{\rm Re}\nolimits} \left\{ {\bar n_{RD}^ *  } \right\}} \right) + \lambda _T \left( {\gamma _{TD}
\hat d_T^{\left( {{\rm{nok}}} \right)}  + 2\sqrt {\gamma _{TD} } d_T {\mathop{\rm Re}\nolimits} \left\{ {\bar n_{TD}^ *  } \right\}} \right)
\vspace*{-10pt}
\end{equation}
\normalsize \hrulefill \vspace*{-10pt}
\end{figure*}
\begin{figure*}[!t]
\setcounter{equation}{8}
\begin{equation} \scriptsize
\label{Eq_9}
\begin{split}
 \mathcal{G}\left( {s,\theta _1 ,\theta _2 } \right) &= {\rm{E}}_{\gamma _{SD} } \left\{ {\exp \left( { - s\gamma _{SD} d_S^2  + s^2 \gamma _{SD} d_S^2 } \right)} \right\}{\rm{E}}_{\gamma _{SR} ,\gamma _{RD} } \left\{ {\exp \left( { - \frac{{\gamma _{SR} }}{{\sin ^2 \left( {\theta _1 } \right)}} - s\min \left\{ {\gamma _{SR} ,\gamma _{RD} } \right\}\hat d_R^{\left( {{\rm{nok}}} \right)}  + s^2 \frac{{\min \left\{ {\gamma _{SR} ,\gamma _{RD} } \right\}}}{{\gamma _{RD} }}d_R^2 } \right)} \right\} \\
  &\times {\rm{E}}_{\gamma _{ST} ,\gamma _{TD} } \left\{ {\exp \left( { - \frac{{\gamma _{ST} }}{{\sin ^2 \left( {\theta _2 } \right)}} - s\min \left\{ {\gamma _{ST} ,\gamma _{TD} } \right\}\hat d_T^{\left( {{\rm{nok}}} \right)}  + s^2 \frac{{\min \left\{ {\gamma _{ST} ,\gamma _{TD} } \right\}}}{{\gamma _{TD} }}d_T^2 } \right)} \right\} \\
 \end{split} \vspace*{-10pt}
\end{equation}
\normalsize \hrulefill \vspace*{-10pt}
\end{figure*}
\begin{figure*}[!t]
\setcounter{equation}{9}
\begin{equation} \scriptsize
\label{Eq_10} \Psi _0 \left( s \right) = \begin{cases}
 \left[ {\bar \gamma _{SD} d_S^2 s\left( {1 - s} \right)} \right]^{ - 1} \hspace{-0.25cm} & d_S  \ne 0 \\
 1 \hspace{-0.25cm} & d_S  = 0 \\
 \end{cases}, \, \Psi _1 \left( {s,\theta _1 } \right) = \begin{cases}
 \left[ {\bar \gamma _{SR} \left( {s\hat d_R^{\left( {{\rm{nok}}} \right)}  + \sin ^{ - 2} \left( {\theta _1 } \right)} \right)} \right]^{ - 1} \hspace{-0.25cm} & d_R  \ne 0 \\
 0 \hspace{-0.25cm} & d_R  = 0 \\
 \end{cases}, \, \Psi _2 \left( {s,\theta _2 } \right) = \begin{cases}
 \left[ {\bar \gamma _{ST} \left( {s\hat d_T^{\left( {{\rm{nok}}} \right)}  + \sin ^{ - 2} \left( {\theta _2 } \right)} \right)} \right]^{ - 1} \hspace{-0.25cm} & d_T  \ne 0 \\
 0 \hspace{-0.25cm} & d_T  = 0 \\
 \end{cases} \vspace*{-5pt}
\end{equation}
\normalsize \hrulefill \vspace*{-5pt}
\end{figure*}
\begin{figure*}[!t]
\setcounter{equation}{11}
\begin{equation} \scriptsize
\label{Eq_12} \mathcal{I}_4 \left( {\hat d_R^{\left( {{\rm{nok}}} \right)} ,\hat d_T^{\left( {{\rm{nok}}} \right)} } \right) = \frac{1}{{2\pi
j}}\int\nolimits_{\delta  - j\infty }^{\delta  + j\infty } {\frac{1}{{s^4 \left( {1 - s} \right)}}\left[ {1 - \left( {1 + s\hat d_R^{\left(
{{\rm{nok}}} \right)} } \right)^{ - {1 \mathord{\left/
 {\vphantom {1 2}} \right.
 \kern-\nulldelimiterspace} 2}} } \right]\left[ {1 - \left( {1 + s\hat d_T^{\left( {{\rm{nok}}} \right)} } \right)^{ - {1 \mathord{\left/
 {\vphantom {1 2}} \right.
 \kern-\nulldelimiterspace} 2}} } \right]ds} \vspace*{-5pt}
\end{equation}
\normalsize \hrulefill \vspace*{-5pt}
\end{figure*}

The next step is the computation of each conditional probability $\Pr \left\{ {\left. {\Delta _{{\bf{c}},{\bf{\tilde c}}} } < 0 \right|\left(
\cdot  \right)} \right\}$. To this end, a closed--form expression of ${\Delta _{{\bf{c}},{\bf{\tilde c}}} }$ is needed. This can be obtained by
substituting $y_{SD}$, $y_{RD}$, and $y_{TD}$ in (\ref{Eq_2}), and through some algebraic manipulations. The final result is as follows:
\setcounter{equation}{4}
\begin{equation} \scriptsize
\label{Eq_5}
\begin{split}
 {\Delta _{{\bf{c}},{\bf{\tilde c}}} }  &= \gamma _{SD} d_S^2  + 2\sqrt {\gamma _{SD} } d_S {\mathop{\rm Re}\nolimits} \left\{ {\bar n_{SD}^ *  } \right\} \\
  &+ \lambda _R \left( {\gamma _{RD} \hat d_R  + 2\sqrt {\gamma _{RD} } d_R {\mathop{\rm Re}\nolimits} \left\{ {\bar n_{RD}^ *  } \right\}} \right) \\
  &+ \lambda _T \left( {\gamma _{TD} \hat d_T  + 2\sqrt {\gamma _{TD} } d_T {\mathop{\rm Re}\nolimits} \left\{ {\bar n_{TD}^ *  } \right\}} \right) \\
 \end{split}
\end{equation}
\noindent where: i) ${\mathop{\rm Re}\nolimits} \left\{  \cdot  \right\}$ is the real part operator; ii) $\left(  \cdot  \right)^ *$ denotes
complex conjugate; iii) $j = \sqrt { - 1}$ is the imaginary unit; iv) $\phi _{XY}$ is the phase of the generic fading gain $h_{XY}$,
\emph{i.e.}, $h_{XY}  = \left| {h_{XY} } \right|\exp \left( {j\phi _{XY} } \right)$; v) $\bar n_{XY}^ *   = {{n_{XY}^ *  \phi _{XY} }
\mathord{\left/ {\vphantom {{n_{XY}^ *  \phi _{XY} } {\sqrt {N_0 } }}} \right. \kern-\nulldelimiterspace} {\sqrt {N_0 } }}$ is the normalized
AWGN for the generic $X$--to--$Y$ link, which has zero mean and unit variance; vi) $d_S  = 2\left( {\tilde c_1  - c_1 } \right) = 2\left(
{\tilde b_S  - b_S } \right)$, $d_R  = 2\left( {\tilde c_2  - c_2 } \right) = 2\left( {\tilde b_S  - b_S } \right)$, $d_T  = 2\left( {\tilde
c_3  - c_3 } \right) = 2\left[ {\left( {\tilde b_S  \oplus \tilde b_T } \right) - \left( {b_S  \oplus b_T } \right)} \right]$; and vii) $\hat
d_R  = 2\left( {1 - 2\hat b_S^{\left( R \right)} } \right)d_R$, $\hat d_T  =  2\left[ {1 - 2\left( {\hat b_S^{\left( T \right)}  \oplus b_T }
\right)} \right]d_T$. Finally, it is worth noticing that the expression given in (\ref{Eq_5}) is useful whichever the conditioning on the bits
estimated at node $R$ and node $T$ are. Only $\hat d_R$ and $\hat d_T$ change for different detection outcomes. To make this aspect more
explicit, we use the notation ($X=R$, $X=T$): i) $\hat d_X  = \hat d_X^{\left( {{\rm{ok}}} \right)}$ if $\hat b_S^{\left( X \right)} = b_S$;
and ii) $\hat d_X  = \hat d_X^{\left( {{\rm{nok}}} \right)}$ if $\hat b_S^{\left( X \right)} \ne b_S$.

To compute $\Pr \left\{ {\left. {\Delta _{{\bf{c}},{\bf{\tilde c}}}  < 0} \right|\left(  \cdot  \right)} \right\}$, we exploit the Laplace
inversion transform method in \cite[Eq. (5)]{Biglieri}:
\setcounter{equation}{5}
\begin{equation} \scriptsize
\label{Eq_6} \Pr \left\{ {\left. {\Delta _{{\bf{c}},{\bf{\tilde c}}}  < 0} \right|\left(  \cdot  \right)} \right\} = \frac{1}{{2\pi
j}}\int\nolimits_{\delta - j\infty }^{\delta  + j\infty } {\frac{{\mathcal{M}_{\Delta _{{\bf{c}},{\bf{\tilde c}}} } \left( {\left. s
\right|\left( \cdot \right)} \right)}}{s}ds}
\end{equation}
\noindent with: i) $\mathcal{M}_{\Delta _{{\bf{c}},{\bf{\tilde c}}} } \left( {\left. s \right|\left(  \cdot  \right)} \right) =
{\rm{E}}_{\left\{ {h_{XD} } \right\},\left\{ {n_{XD} } \right\}} \left\{ {\left. {\exp \left( { - s\Delta _{{\bf{c}},{\bf{\tilde c}}} }
\right)} \right|\left( \cdot  \right)} \right\}$ being the (two--sided) Moment Generating Function (MGF) of the conditional RV ${\Delta
_{{\bf{c}},{\bf{\tilde c}}} }$. The average is computed over fading gains and AWGN of all the links $X$--to--$D$ for $X = \left\{ {S,R,T}
\right\}$; and ii) $\delta$ being a real number such that the contour path of integration is in the region of convergence of
${\mathcal{M}_{\Delta _{{\bf{c}},{\bf{\tilde c}}} } \left( {\left. \cdot \right| \cdot } \right)}$.

Then, ${\rm{APEP}}\left( {{\bf{c}} \to {\bf{\tilde c}}} \right)$ can be obtained by substituting (\ref{Eq_6}) in (\ref{Eq_4}), by computing the
expectation over fading statistics, AWGN, and by solving the inverse Laplace transform. In particular, since in this paper we are interested in
high--SNR analysis, \emph{i.e.}, ${{E_m } \mathord{\left/ {\vphantom {{E_m } {N_0  \to \infty }}} \right. \kern-\nulldelimiterspace} {N_0 \to
\infty }}$, an asymptotic expression of the MGF in (\ref{Eq_6}) is needed \cite[Eq. (12)]{Biglieri}. Due to space constraints, in this paper we
cannot provide all the details of the derivation. As an illustrative example, we provide a brief description of the main steps behind the
computation of one addend in (\ref{Eq_4}). In particular, we focus our attention on the fourth addend in (\ref{Eq_4}), which is denoted by
${\rm{APEP}}^{\left( {\rm{4}} \right)} \left( {{\bf{c}} \to {\bf{\tilde c}}} \right)$. The reason is that this term is the most complicated to
be computed.

${\rm{APEP}}^{\left( {\rm{4}} \right)} \left( {{\bf{c}} \to {\bf{\tilde c}}} \right)$ in (\ref{Eq_4}) can be written as shown in (\ref{Eq_7})
on top of the next page, where: i) $\mathcal{F}\left( \cdot, \cdot \right)$ is defined in (\ref{Eq_8}) on top of the next page; ii) $\mathop  =
\limits^{\left( a \right)}$ is obtained by using the Craig's representation of the Q--function \cite{Simon}; iii) $\mathop  = \limits^{\left( b
\right)}$ is obtained by averaging over the AWGN with $\mathcal{G}\left( {\cdot,\cdot,\cdot} \right)$ being defined in (\ref{Eq_9}) on top of
the next page; and iv) $\mathop  = \limits^{\left( c \right)}$ is obtained by averaging over channel fading and using some simplifications that
hold for high--SNR. In particular, ${\Psi _0 \left( \cdot \right)}$, ${\Psi _1 \left( {\cdot,\cdot } \right)}$, and ${\Psi _2 \left(
{\cdot,\cdot } \right)}$ are defined in (\ref{Eq_10}) on top of the next page, where $\bar \gamma _{XY}   = 2\sigma _{XY}^2 \left( {E_m /N_0 }
\right)$ for the generic pair of nodes $X$ and $Y$. Note that, for $X=R$ and $X=T$, $d_X \ne 0 \Leftrightarrow \hat d_X \ne 0$.

Let us consider the most general case with $d_S  \ne 0$, $d_R  \ne 0$, and $d_T  \ne 0$. Both integrals in the brackets in (\ref{Eq_10}) can be
computed in closed--form with the help of \cite[Eq. (5A.9)]{Simon}. Thus, ${\rm{APEP}}^{\left( {\rm{4}} \right)} \left( {{\bf{c}} \to
{\bf{\tilde c}}} \right)$ simplifies as follows:
\setcounter{equation}{10}
\begin{equation} \scriptsize
\label{Eq_11} {\rm{APEP}}^{\left( {\rm{4}} \right)} \left( {{\bf{c}} \to {\bf{\tilde c}}} \right) = \frac{{\mathcal{I}_4 \left( {\hat
d_R^{\left( {{\rm{nok}}} \right)} ,\hat d_T^{\left( {{\rm{nok}}} \right)} } \right)}}{{4\bar \gamma _{SD} \bar \gamma _{SR} \bar \gamma _{ST}
d_S^2 \hat d_R^{\left( {{\rm{nok}}} \right)} \hat d_T^{\left( {{\rm{nok}}} \right)} }}
\end{equation}
\noindent where $\mathcal{I}_4 \left( { \cdot , \cdot } \right)$ is defined in (\ref{Eq_12}) on top of the next page.

Some important considerations are worth being made about ${\rm{APEP}}^{\left( {\rm{4}} \right)} \left( {{\bf{c}} \to {\bf{\tilde c}}} \right)$
in (\ref{Eq_11}). First, we notice that the asymptotic behavior of the APEP is clearly shown, and, for the considered case study, a diversity
order equal to three is obtained \cite{Ribeiro}. Second, the integral $\mathcal{I}_4 \left( { \cdot , \cdot } \right)$ can be computed, either
analytically or numerically, by using one of the many methods described in \cite{Biglieri}. Finally, we would like to mention that the case
study investigated in this section, \emph{i.e.}, ${\rm{APEP}}^{\left( {\rm{4}} \right)} \left( {{\bf{c}} \to {\bf{\tilde c}}} \right)$, is the
most complicated addend, as it is the only term involving the product of two Q--functions. All the other cases are much simpler to be computed,
and all integrals similar to $\mathcal{I}_4 \left( { \cdot , \cdot } \right)$ in (\ref{Eq_12}) can be computed in closed--form by using the
method of residues \cite[Eq. (6)]{Biglieri}. The details of the derivation are omitted, but final results are summarized and discussed in
Section \ref{Comparison}.
\section{Performance Comparison: Is NC Useful?} \label{Comparison}
The aim of this section is to compare the performance of the different scenarios and network topologies shown in Fig. \ref{Fig_1} and Fig.
\ref{Fig_2}. For all cases of interest, the methodology described in Section \ref{Framework} is used to compute the ABEP. In particular,
(\ref{Eq_3}) is applied for all possible codewords of the codebook. The final results are summarized in Table \ref{Tab_1}, by assuming, for a
fair comparison, the total energy constraint mentioned in Section \ref{SignalModel}. Furthermore, since we are interested in high--SNR
analysis, Table \ref{Tab_1} shows only the dominant terms in (\ref{Eq_3}), \emph{i.e.}, those APEPs having the slowest decaying behavior as a
function of ${{E_m } \mathord{\left/ {\vphantom {{E_m } {N_0 \to \infty }}} \right. \kern-\nulldelimiterspace} {N_0  \to \infty }}$
\cite{Iezzi_ICC2011}. In fact, these terms determine both diversity and coding gain. The accuracy of the frameworks shown in Table \ref{Tab_1}
is validated in Section \ref{Results} through Monte Carlo simulations.

Important considerations can be drawn from our analysis. Let us consider the 3--node network topology. The ABEP of Scenario (b) shows that node
$S$ can exploit distributed diversity to improve the diversity gain, but the price to pay is a performance degradation for node $R$, whose ABEP
is worse than in the non--cooperative case, \emph{i.e.}, Scenario (a). Very interestingly, we notice that the network--coded scenario,
\emph{i.e.}, Scenario (c), is the worst one in terms of performance. Node $S$ has no gain from cooperation, and the diversity order is equal to
one. Furthermore, and very surprisingly, node $S$ has the same ABEP as in the non--cooperative case. In other words, there is neither power nor
diversity gain. As far as node $R$ is concerned, the situation is even worse: the ABEP is worse than the non--cooperative case. Also, we notice
that this performance penalty depends only in part on decoding errors on the $S$--to--$R$ link. In fact, even assuming $\bar \gamma _{SR}  \to
\infty$, \emph{i.e.}, no decoding errors at node $R$, the ABEP is worse because of performing NC. In conclusion, unlike \cite{Nasri},
\cite{Iezzi_ICC2011}--\cite{Iezzi_TIT2011} where it shown that NC is beneficial in cooperative networks when some nodes act only as relays
and have no data to transmit, Table \ref{Tab_1} points out that, if the relay nodes have their own data to transmit, NC introduces no gain when
compared to the non--cooperative scenario, and, in some cases, NC might also be harmful. To the best of the authors knowledge, this important
behavior has never been reported in the open technical literature \cite{Zorzi}. Similar comments apply to the 4--node network topology. In
particular, we notice that node $S$ has a diversity order that depends on the number of relay nodes that do not perform NC but just forward the
received packets.

Finally, we would like to emphasize that, unlike state--of--the--art performance analysis of cooperative networks (see \cite{Cano},
\cite{Nasri}, \cite{Iezzi_GLOBECOM2011} for further comments), our analysis encompasses a very accurate estimation of the coding gain. This is
instrumental to clearly assess diversity and coding trade--off summarized in Table \ref{Tab_1}.
\begin{table*}[!t] \scriptsize
\renewcommand{\arraystretch}{1.75}
\caption{\scriptsize ABEP for high--SNR ($k_1 \approx 0.4853$ is obtained by computing terms like $I_4 \left( {\hat d_R^{\left( {{\rm{nok}}}
\right)} ,\hat d_T^{\left( {{\rm{nok}}} \right)} } \right)$ in (\ref{Eq_11}) by using \cite[Eq. (10)]{Biglieri}, and $k_2  = \left( {525 +
11\sqrt 5 } \right)/800$). \vspace{-0.5cm}} \label{Tab_1}
\begin{center}
\begin{tabular}{|c||c|c|c|}
\hline
 & ${\rm{ABEP}}_S$ & ${\rm{ABEP}}_R$ & ${\rm{ABEP}}_T$ \\
\hline
3--Node Network (a) & $\left( {1/4} \right)\bar \gamma _{SD}^{ - 1}$ & $\left( {1/4} \right)\bar \gamma _{RD}^{ - 1}$ & -- \\
\hline
3--Node Network (b) & $\left( {3/8} \right)\bar \gamma _{SD}^{ - 1} \bar \gamma _{RD}^{ - 1}  + \left[ {\left( {45 + \sqrt 5 } \right)/160} \right]\bar \gamma_{SD}^{ - 1} \bar \gamma _{SR}^{ - 1}$ & $\left( {1/2} \right)\bar \gamma _{RD}^{ - 1}$ & -- \\
\hline
3--Node Network (c) & $\left( {1/4} \right)\bar \gamma _{SD}^{ - 1}$ & $\left( {1/4} \right)\bar \gamma _{SD}^{ - 1}  + \left( {1/4} \right)\bar \gamma _{SR}^{ - 1}  + \left( {1/4} \right)\bar \gamma _{RD}^{ - 1}$ & -- \\
\hline \hline
4--Node Network (a) & $\left( {1/4} \right)\bar \gamma _{SD}^{ - 1}$ & $\left( {1/4} \right)\bar \gamma _{RD}^{ - 1}$ & $\left( {1/4} \right)\bar \gamma _{TD}^{ - 1}$ \\
\hline 4--Node Network (b) & $\begin{array}{l}
\left( {5/8} \right)\bar \gamma _{SD}^{ - 1} \bar \gamma _{RD}^{ - 1} \bar \gamma _{TD}^{ - 1}  + k_1 \bar \gamma _{SD}^{ - 1} \bar \gamma _{SR}^{ - 1} \bar \gamma _{ST}^{ - 1}  \\
+ k_2 \bar \gamma _{SD}^{ - 1} \bar \gamma _{SR}^{ - 1} \bar \gamma _{TD}^{ - 1}  + k_2 \bar \gamma _{SD}^{ - 1} \bar \gamma _{ST}^{ - 1} \bar \gamma _{RD}^{ - 1}  \\
\end{array}$ & $\left( {1/2} \right)\bar \gamma _{RD}^{ - 1}$ & $\left( {1/2} \right)\bar \gamma _{TD}^{ - 1}$ \\
\hline
4--Node Network (c) & $\left( {1/4} \right)\bar \gamma _{SD}^{ - 1}$ & $\left( {1/4} \right)\bar \gamma _{SD}^{ - 1}  + \left( {1/4} \right)\bar \gamma _{SR}^{ - 1}  + \left( {1/4} \right)\bar \gamma _{RD}^{ - 1}$ & $\left( {1/4} \right)\bar \gamma _{SD}^{ - 1}  + \left( {1/4} \right)\bar \gamma _{ST}^{ - 1}  + \left( {1/4} \right)\bar \gamma _{TD}^{ - 1}$ \\
\hline
4--Node Network (d) & $\left( {3/8} \right)\bar \gamma _{SD}^{ - 1} \bar \gamma _{RD}^{ - 1}  + \left[ {\left( {45 + \sqrt 5 } \right)/160} \right]\bar \gamma _{SD}^{ - 1} \bar \gamma _{SR}^{ - 1}$ & $\left( {1/2} \right)\bar \gamma _{RD}^{ - 1}$ & $\left( {1/4} \right)\bar \gamma _{ST}^{ - 1}  + \left( {1/4} \right)\bar \gamma _{TD}^{ - 1}$ \\
\hline
\end{tabular}
\end{center} \vspace{-0.5cm}
\end{table*}
\section{Numerical and Simulation Results} \label{Results}
\begin{figure}[!t]
\centering
\includegraphics [width=\columnwidth] {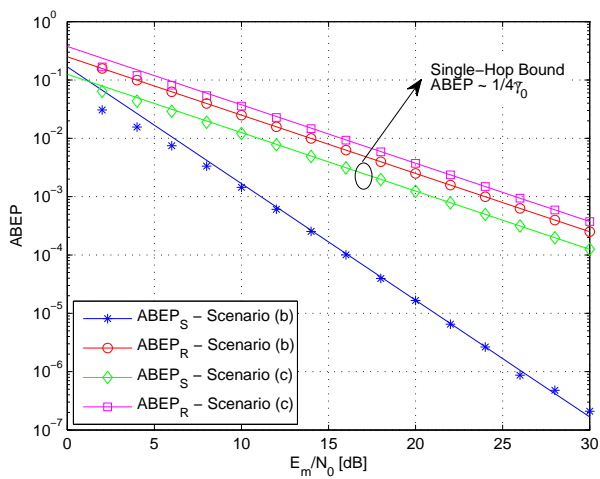}
\vspace{-0.75cm} \caption{\scriptsize ABEP against $E_m/N_0$ for the 3--node network topology in Fig. \ref{Fig_1}. Solid lines show the
analytical framework and markers show Monte Carlo simulations. Setup: i) i.i.d. fading with $\sigma _0^2 = 1$; and ii) $\bar \gamma _0 =
2\sigma _0^2 \left( {E_m /N_0 } \right)$. ${\rm{ABEP}}_S$ and ${\rm{ABEP}}_R$ of Scenario (a) are given by the single--hop bound.}
\label{Fig_3}
\end{figure}

\begin{figure}[!t]
\centering
\includegraphics [width=\columnwidth] {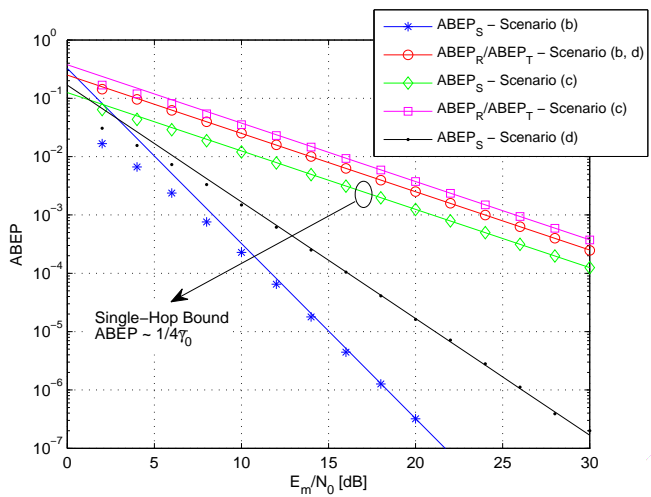}
\vspace{-0.75cm} \caption{\scriptsize ABEP against $E_m/N_0$ for the 4--node network topology in Fig. \ref{Fig_2}. Solid lines show the
analytical framework and markers show Monte Carlo simulations. Setup: i) i.i.d. fading with $\sigma _0^2 = 1$; and ii) $\bar \gamma _0 =
2\sigma _0^2 \left( {E_m /N_0 } \right)$. ${\rm{ABEP}}_S$, ${\rm{ABEP}}_R$, and ${\rm{ABEP}}_T$ of Scenario (a) are given by the single--hop
bound.} \label{Fig_4} \vspace{-0.25cm}
\end{figure}

In this section, we compare the frameworks summarized in Table \ref{Tab_1} with Monte Carlo simulations. More specifically, simulation results
are obtained through a brute force implementation of (\ref{Eq_2}). Some selected curves are shown in Fig. \ref{Fig_3} and Fig. \ref{Fig_4} for
the 3--node and 4--node scenario, respectively. For simplicity, but without loss of generality, i.i.d. fading is considered. We
can see that the framework in Table \ref{Tab_1} closely overlaps with Monte Carlo simulations for high--SNR. This confirms the accuracy of the
analytical derivation in Section \ref{Framework}, and the theoretical findings Section \ref{Comparison}.
\section{Conclusion} \label{Conclusion}
In this paper, we have studied the performance of network--coded cooperative wireless networks with practical communication constraints. A
general framework has been proposed, which can capture diversity and coding gain, and provides insightful information about the performance of
the system, along with the tradeoff and the interplay of cooperation and NC. Unlike common belief, our analysis has clearly shown that using NC
might be harmful for the system. In fact, we have shown that the diversity order is determined only by those nodes that act as repeaters and do
not network--code their own data to the received packets. These results and conclusions are valid for binary modulation and binary NC. Current
research activity is now concerned with the investigation of wireless networks with non--binary modulation and non--binary NC.
\section*{Acknowledgment} \footnotesize
This work is supported, in part, by the research projects ``GREENET'' (PITN--GA--2010--264759), ``WSN4QoL'' (IAPP--GA--2011--286047), and the
Lifelong Learning Programme (LLP) -- ERASMUS Placement.

\begin{thebibliography}{99}
%
\bibitem{Nosratinia} A. Nosratinia, T. E. Hunter, and A. Hedayat,
              ``Cooperative communications in wireless networks'',
             \emph{IEEE Commun. Mag.},
             vol. 42, no. 10, pp. 74--80, Oct. 2004.
%
\bibitem{Laneman} J. N. Laneman, D. Tse, and G. Wornell,
              ``Cooperative diversity in wireless networks: Efficient protocols and outage behavior'',
             \emph{IEEE Trans. Inform. Theory},
             vol. 50, no. 12, pp. 3062--3080, Dec. 2004.
%
\bibitem{Krikidis} Z. Ding \emph{et al.},
              ``On combating the half--duplex constraint in modern cooperative networks: Protocols and techniques'',
             \emph{IEEE Wireless Commun. Mag.},
             Apr. 2011. [Online]. Available: http://www.staff.ncl.ac.uk/z.ding/WC\_magazine.pdf.
%
\bibitem{Zorzi} F. Rossetto and M. Zorzi,
              ``Mixing network coding and cooperation for reliable wireless communications'',
             \emph{IEEE Wireless Commun. Mag.},
             vol. 18, no. 1, pp. 15--21, Feb. 2011.

\bibitem{Ahlswede} R. Ahlswede \emph{et al.},
             ``Network information flow'',
             \emph{IEEE Trans. Inform. Theory},
             vol. 46, no. 4, pp. 1204--1216, July 2000.
%
\bibitem{Gerla} J.--S. Park \emph{et al.},
              ``Codecast: A network--coding--based ad hoc multicast protocol'',
             \emph{Wireless Commun.},
             vol. 13, no. 5, pp. 76--81, Oct. 2006.
%
\bibitem{Katti_PhD} S. Katti,
              ``Network coded wireless architecture'',
             \emph{Ph.D. Dissertation},
             Massachusetts Institute of Technology, Sep. 2008.
%
\bibitem{Ribeiro} A. Ribeiro, X. Cai, and G. Giannakis,
             ``Symbol error probabilities for general cooperative links'',
             \emph{IEEE Trans. Wireless Commun.},
             vol. 4, no. 3, pp. 1264--1273, May 2005.
%
\bibitem{MDR_TCOMSep2009} M. Di Renzo, F. Graziosi, and F. Santucci,
             ``A unified framework for performance analysis of CSI--assisted cooperative communications over fading channels'',
             \emph{IEEE Trans. Commun.},
             pp. 2552--2557, Sep. 2009.
%
\bibitem{MDR_TWCOct2009} ---,
             ``A comprehensive framework for performance analysis of dual--hop cooperative wireless systems with fixed--gain relays over generalized fading channels'',
             \emph{IEEE Trans. Wireless Commun.},
             vol. 8, Oct. 2009.
%
\bibitem{MDR_TCOMFeb2010} ---,
             ``A comprehensive framework for performance analysis of cooperative multi--hop wireless systems over log--normal fading channels'',
             \emph{IEEE Trans. Commun.},
             vol. 58, no. 2, pp. 531--544, Feb. 2010.
%
\bibitem{MDR_Springer2010} M. Di Renzo \emph{et al.},
             ``Robust wireless network coding -- An overview'',
             \emph{Springer Lecture Notes},
             LNICST 45, pp. 685--698, 2010.
%
\bibitem{Nguyen2010} S. L. H. Nguyen \emph{et al.},
             ``Mitigating error propagation in two--way relay channels with network coding'',
             \emph{IEEE Trans. Wireless Commun.},
             vol. 9, pp. 3380--3390, Nov. 2010.
%
\bibitem{AlHabian2011} G. Al--Habian \emph{et al.},
             ``Threshold--based relaying in coded cooperative networks'',
             \emph{IEEE Trans. Veh. Technol.},
             vol. 60, pp. 123--135, Jan. 2011.
%
\bibitem{Cano} A. Cano \emph{et al.},
             ``Link--adaptive distributed coding for multi--source cooperation'',
             \emph{EURASIP J. Adv. Signal Process.},
             vol. 2008, Jan. 2008.
%
\bibitem{Nasri} A. Nasri, R. Schober, and M. Uysal,
             ``Error rate performance of network--coded cooperative diversity systems'',
             \emph{IEEE Global Commun. Conf.},
             pp. 1–6, Dec. 2010.
%
\bibitem{RayLiu} H.–Q. Lai and K. J. Ray Liu,
             ``Space--time network coding'',
             \emph{IEEE Trans. Signal Process.},
             vol. 59, no. 4, pp. 1706--1718, Apr. 2011.
%
\bibitem{Perez--Neira} G. Li \emph{et al.},
             ``High--throughput multi--source cooperation via complex--field network coding'',
             \emph{IEEE Trans. Wireless Commun.},
             vol. 10, no. 5, pp. 1606--1617, May 2011.
%
\bibitem{Iezzi_ICC2011} M. Iezzi, M. Di Renzo, and F. Graziosi,
             ``Network code design from unequal error protection coding: Channel--aware receiver design and diversity analysis'',
             \emph{IEEE Int. Commun. Conf.},
             pp. 1--6, June 2011.
%
\bibitem{Iezzi_GLOBECOM2011} ---,
             ``Closed--form error probability of network--coded cooperative wireless networks with channel--aware detectors'',
             \emph{IEEE Global Commun. Conf.},
             pp. 1--6, Dec. 2011.
%
\bibitem{Iezzi_TIT2011} ---,
             ``Diversity and coding gain of multi--source multi--relay cooperative wireless networks with binary network coding'',
             pp. 1--56, Sep. 2011, submitted. [Online]. Available: http://arxiv.org/pdf/1109.4599v1.pdf.
%
\bibitem{GiannakisLaneman} T. Wang \emph{et al.},
             ``High--performance cooperative demodulation with decode--and--forward relays'',
             \emph{IEEE Trans. Commun.},
             vol. 55, no. 7, pp. 1427--1438, Jul. 2007.
%
\bibitem{Proakis} J. J. Proakis,
              \emph{Digital Communications},
             McGraw--Hill, 4th ed., 2000.
%
\bibitem{Biglieri} E. Biglieri \emph{et al.},
             ``Computing error probabilities over fading channels: A unified approach'',
             \emph{European Trans. Telecommun.},
             vol. 9, no. 1, pp. 15--25,  Jan.--Feb. 1998.
%
\bibitem{Simon} M. K. Simon and M.--S. Alouini,
              \emph{Digital Communication over Fading Channels},
             John Wiley $\&$ Sons, Inc., 1st ed., 2000.
%
%
\end{thebibliography}
\end{document}